# Developing a Conceptual Tribal Crash Safety Dashboard: Data-Driven Strategies for Identifying High-Risk Areas and Enhancing Tribal Safety Programs

Tianyi Chen[a], Haotian Shi*[a], Steven T. Parker[a], Glenn Vorhes[a], David A Noyce[a], Bin Ran[a]

[a]*Department of Civil and Environmental Engineering, University of Wisconsin-Madison, Madison, United States;*

*Corresponding author: Haotian Shi (email address: hshi84@wisc.edu)

# Developing a Conceptual Tribal Crash Safety Dashboard: Data-Driven Strategies for Identifying High-Risk Areas and Enhancing Tribal Safety Programs

Tribal lands in the United States have consistently exhibited higher crash rates and injury severities compared to other regions. To address this issue, effective data-driven safety analysis methods are essential for resource allocation and tribal safety program development. This study outlines the minimum data requirements and presents a generic tribal crash dashboard prototype to enable feasible and reproducible tribal crash analysis. The dashboard offers statistical performance measurement techniques, tracks tribal safety trends using various indicators, and updates in with incoming data, making it adaptable for any state. Specifically, the dashboard's capabilities are showcased using Wisconsin tribal crash data, locating high-risk tribal land areas while analyzing and comparing statewide crashes and tribal crashes based on severity. The results reveal that tribal land roads, particularly rural ones, are more dangerous than those in other Wisconsin areas. The dashboard also examines crash types to uncover the causes and insights behind Wisconsin tribal crashes. Demonstrating this dashboard concept reveals its potential to facilitate a more intuitive understanding of tribal crashes from a statistical standpoint, thereby enabling more effective applications of available data sources and the development of targeted safety countermeasures.

# 1. Introduction

Tribal lands have historically exhibited elevated crash rates and injury severities compared to other regions throughout the United States (e.g., Opus International Consultants & Ceifetz, 2012; Bailey & Huft, 2008; Shinstine & Ksaibati, 2013). To mitigate the consequences of these crashes, it is imperative to establish effective transportation safety programs supported by agencies such as the Federal Highway Administration (FHWA) (Highway Safety Improvement Program [HSIP], 2022), Bureau of Indian Affairs (BIA) (Indian Highway Safety Program, 2022), Tribal Transportation Program Safety Fund (Tribal Transportation Program Safety Fund [TTPSF], 2022), and National Highway Traffic Safety Administration (NHTSA) (Traffic Records Coordinating Committee, 2022). Furthermore, a cohesive program for tribal transportation facilities, initiated with MAP-21 (MAP-21 - Moving Ahead for Progress in the 21st Century, 2022), has been consistently expanded through subsequent legislation, including the recent Bipartisan Infrastructure Law, to generate funding opportunities for tribal roads and transit systems (U.S. Department of Transportation, 2022).

These opportunities rely on data-driven methodologies, necessitating the implementation of efficient performance measurement strategies based on the analysis of tribal crash data. Leveraging the resources and opportunities provided by the agencies above, various transportation organizations have developed tribal safety programs that incorporate data-driven approaches with differing levels of success.

The Wyoming Technology Transfer Center (Shinstine & Ksaibati, 2013) has developed a system for identifying high-risk zones on tribal roads using a five-step approach, comprising (i) crash analysis, (ii) Level I field evaluation, (iii) combined ranking, (iv) Level II field evaluation, and (v) benefit-cost analysis. This cost-effective safety enhancement technique effectively detects high-risk areas on low-volume roadways and has the potential to reduce the elevated crash rates in tribal lands. By examining crash data trends, the five-step process mitigates the negative impact of underreporting and facilitates more informed decisions regarding safety-improvement measures. Nevertheless, the complexity of this methodology's framework and steps, such as manual crash report transmission and evaluation by a panel of tribal experts, poses challenges for automated processing.

The North Dakota Department of Transportation (NDDOT) (Vachal, 2022) recognized that tribal lands typically lack a comprehensive crash reporting system (CRS) and established a framework to improve the CRS for tribal communities in North Dakota. Through collaboration with other stakeholders and the tribes' commitment, the study demonstrated the feasibility of electronically documenting motor vehicle (MV) crash occurrence data. Nonetheless, it does not systematically perform high-risk crash analyses nor delve deeper into the investigation of specific crash types (e.g., alcohol-related, pedestrian, animal).

To reduce the frequency and severity of crash-related injuries, Northern Arizona University (James & Russo, 2019) conducted an in-depth investigation of crash severity characteristics across five major tribal lands in Arizona. The study identified several factors related to the individual, vehicle, roadway, and environment influencing crash severity. This report can aid law enforcement and transportation agencies in formulating strategies to decrease tribal crashes and enhance safety within tribes. Additionally, incorporating the examined factors into tribal crash reporting could improve the data quality used for post-safety investigations. However, the research did not consider certain tribal factors impacting injury severity, such as data on emergency medical service response times.

The Minnesota Roadway Safety Institute explored the potential of Geographic Information Systems (GIS) advancements to enhance transportation safety data (Horan et al., 2018). It developed several innovative GIS applications tailored for tribal use, including one that examined hotspots, road segments, and surrounding areas. Additionally, the study established a data framework for evaluating tribal community, governance, and data management, which could be employed to oversee and refine tribal crash reporting. Based on this architecture, the GIS could readily automate hotspot analysis and other tribal safety evaluations. However, despite comprehensive research, the project did not consider detailed factors such as exposure, behavioral aspects, and physical roadway design in its safety analysis.

The Wisconsin Department of Transportation (WisDOT) collects and evaluates data on the recording and handling of tribal traffic crashes by law enforcement agencies (Redinger et al., 2012). Despite significant findings and recommendations for enhancing crash reporting, the study did not thoroughly examine detailed crash attributes (e.g., person-related attributes; crash types). Additionally, WisDOT conducted an extensive tribal crash review in 2012 for the seven largest tribal territories in Wisconsin (Opus International Consultants & Ceifetz, 2012). This study encompassed the process of filing a crash report, interacting with enforcement agencies, and conducting a comprehensive examination of incidents to identify potential avenues for improving tribal safety. However, manual data processing and subpar tribal data quality resulted in an inefficient and inadequate crash analysis. Chen et al. systematically performed quantitative data analysis to verify the effectiveness of tribal crash data attributes based on the Wisconsin crash report. Nevertheless, more in-depth tribal crash analyses, such as hotspot analysis, have not been conducted.

These programs demonstrate that high-quality, spatially referenced crash data is essential for facilitating effective tribal crash analysis. However, the current studies and projects exhibit three notable limitations. First, the majority of tribal crash data systems and analysis methods lack sufficient detail and clarity, making it difficult to reproduce the procedures. Second, few studies investigate high-risk locations within tribal lands and the underlying reasons behind tribal crashes. Identifying high-risk locations and insights are crucial for addressing safety concerns. Third, tribal data analysis is often hampered by data availability and quality issues. It is necessary to explore which types of analyses can be effectively conducted given the minimum baseline set of tribal crash data. Motivated by these gaps, there remains an opportunity to develop a data-driven and reproducible approach for states to identify issues, conduct more in-depth analyses, and implement informed countermeasures based on the analysis results.

Online dashboards have become an increasingly popular method for tracking, analyzing, and visualizing crash data performance indicators, as demonstrated by the development of such tools by various agencies, including the Denver government (Crash Data Dashboard [Denver], 2022), Colorado Department of Transportation (Colorado Crash Data Dashboard, 2022), Houston-Galveston Area Council (Regional Crash Data, 2022), and Orlando Metroplan (Crash Data Dashboard [Metroplan Orlando], 2022). Dashboards effectively display and monitor statistical safety trends concerning diverse indicators (e.g., locations, crash types, crash density, and person-related attributes), provide performance measurement approaches for crash analysis, and can be systematically updated as new source data emerges. These merits of the dashboard concept can be harnessed and integrated with tribal crash systems to facilitate effective tribal crash analysis and bolster tribal safety programs.

Despite the benefits of using dashboards for crash analysis, some limitations remain when conducting tribal crash analysis. Currently, crash dashboards typically focus on directly displaying crash attributes or indicators, which may not yield intuitive insights or expose underlying issues associated with crash attributes. To gain a more comprehensive understanding of tribal land safety performance, comparative analyses can be employed, such as contrasting tribal lands with statewide areas, individual tribes with tribal averages, and different road types. By integrating such comparative analyses into the tribal crash dashboard concept, the dashboard's capabilities can be further enhanced.

Considering the limitations mentioned above, this paper aims to develop an efficient method for resource allocation and rational improvement of traffic safety in tribal lands. The objectives of this study are to (i) identify the minimum data requirements necessary for conducting data analysis; (ii) develop and demonstrate a generic dashboard of analysis techniques based on the available data, which analyzes and identifies high-risk roadways and crash hotspots in tribal lands, serving as a starting point for more in-depth crash analysis; and (iii) carry out comparative analyses for tribal lands.

Specifically, this study creates a generic and reproducible crash dashboard prototype. To demonstrate its analyzing capabilities, this paper conducts a systematic Wisconsin tribal crash analysis using the dashboard as an example. The dashboard is employed to identify high-risk tribal lands in Wisconsin through a comparative analysis with statewide crash indicators. Enabled by the dashboard's analysis, we can obtain a more intuitive understanding of tribal crashes from a statistical standpoint, capitalize on available data sources, and more effectively devise safety countermeasures. The tribal safety dashboard concept presented in this paper also offers a template for other states to develop similar capabilities, along with a demonstration of what can be accomplished based on a minimal set of tribal crash data elements.

In summary, the primary contribution of this work are as follows: (i) the provision of a generic and adaptable tribal crash dashboard framework for tribal safety analysis based on limited data sources; (ii) the integration of ArcGIS capabilities and comparative analysis techniques, facilitating a more thorough examination of high-risk roadways and crash hotspots within tribal lands; (iii) the investigation of tribal crash safety in Wisconsin using the proposed framework, promoting the implementation of safety countermeasures in Wisconsin tribal areas.

**2. Wisconsin tribes and crash reporting**

Eleven federally recognized tribes are located in Wisconsin, as shown in **Figure 1**. Based on existing agreements, ten of these eleven tribal law enforcement agencies electronically report motor vehicle crashes to the State of Wisconsin using the standard Wisconsin DT4000 police crash report provided by WisDOT and the associated Wisconsin Badger Traffic and Criminal Software (TraCS) system (TraCS, 2022). Crash reports submitted through Badger TraCS are automatically stored in the Wisconsin Crash Database, where they are typically available for analysis and reporting the following day. The Menominee Tribe is the only exception, as it has its own crash reporting system and generally only reports fatal crashes to the State.

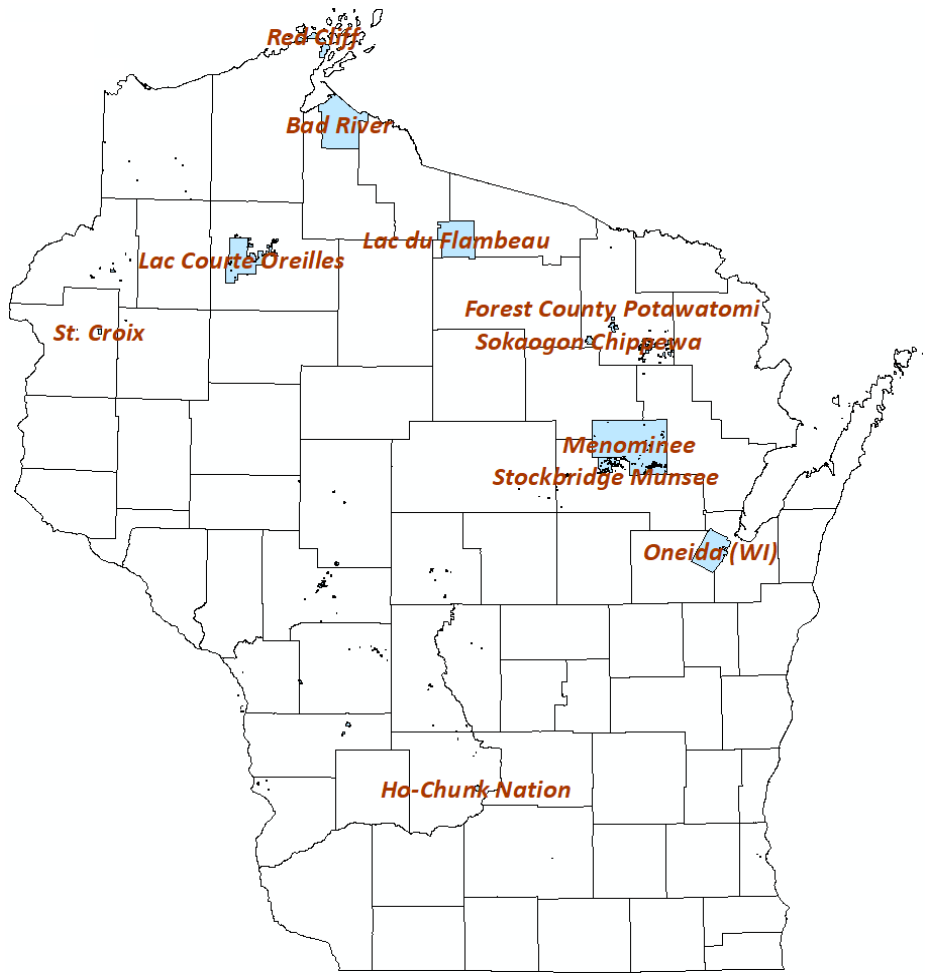

Figure 1. Location of 11 tribal lands in Wisconsin.

The DT4000 crash report form, introduced in 2017, is highly compliant with the NHTSA's Model Minimum Uniform Crash Criteria (MMUCC) 5th edition (National Highway Traffic Safety Administration [NHTSA], 2017) and implements several significant enhancements compared to the previous crash report. Two major improvements relevant to this study include the addition of mapped geographic coordinates for crash locations, which substantially increase the direction and efficiency of post-processing analysis. Another improvement involves the inclusion of tribal-related elements as local extensions to the MMUCC. Specifically, the tribal-related elements are reflected in crash location, jurisdiction, and law enforcement agency type. The detailed changes are presented in **Table 1**.

Table 1. Tribal-related Crash Classification Element Summary in DT4000.

| DT4000 Tribal-related Crash Elements | | |
|---|---|---|
| Element Name | Element Code | Attribute Value Example |
| Crash Classification - Location | CRSHLOC | Public Property, Private Property, Tribal Land |
| Crash Classification - Jurisdiction | CRSHJUR | No Special Jurisdiction, College/University Campus, Military, Notional Park Service, Indian Reservation/Trust, etc. |
| Enforcement Agency Type | AGCYTYPE | State Patrol, County Sheriff, City Police, Tribal, etc. |
| Tribal Code | TRBCODE | "Wisconsin Tribal Codes" |
| | TRBNAME | "Wisconsin Tribal Names" |

With the enhancements mentioned above, the current DT4000 dataset offers comprehensive features for identifying tribal crashes. Additionally, a previous study (*14*) has systematically assessed and validated the effectiveness and quality of the updated tribal-related elements. Although the validation demonstrates that these modifications aid in the rapid identification of tribal crashes, the study also highlights several remaining shortcomings and puts forth suggestions for further refinement.

For instance, it is proposed that tribal-related attributes in Crash Classification - location and - jurisdiction be extracted into a new, independent subfield. This would enable tracking crashes within tribal lands regardless of the public or private roadway type, thus eliminating confusion and inconsistent reporting of this element. It is also recommended that BIA road attributes be incorporated into the roadway functional class to better identify crashes occurring on BIA roads. However, these recommendations have yet to be implemented. To explore the potential benefits of implementing these suggestions, this study manually processed the data and developed a prototype dashboard to safely and flexibly achieve the project objectives.

## 3. Tribal safety dashboard concept overview

The primary objective of the tribal safety dashboard is to create a versatile analysis and visualization tool that meets the minimum data requirements for (i) monitoring statewide tribal crashes, (ii) intuitively identifying issues, and (iii) facilitating in-depth crash analysis. Although the dashboard's focus lies on tribal crashes, it will still reference statewide crash statistics. Consequently, the subsequent sections discuss and emphasize the potential representative elements of the tribal safety dashboard, utilizing the Wisconsin crash database as an example. In particular, the dashboard connects to the DT4000 database, which encompasses a comprehensive record of all reported crashes in Wisconsin. Each crash adheres to the MMUCC data model and is classified according to crash-level, vehicle-level, and person-level elements and attributes.

To obtain a clearer comprehension of the disparities in tribal crashes, this study examines and contrasts crashes occurring throughout Wisconsin and within its tribal lands. **Table 2** presents a summary of the statistics related to the severity levels of statewide crashes and those taking place on tribal lands. It is important to note that the injury severity level adheres to the MMUCC code definitions: fatal injury (K), suspected serious (incapacitating) injury (A), suspected minor (non-incapacitating) injury (B), possible injury (C), and no apparent injury (O) (*16*). This study specifically concentrates on "fatal and serious injuries," encompassing "KA" crashes, as well as a broader dataset that includes minor injury types as "KAB," offering a more substantial data sample for analysis.

The final dataset for 2017-2021 comprises a total of 672,363 crashes statewide, of which 3,396 (0.5%) occurred within tribal lands. The distribution of crash severity among these 3,396 crashes included 20 K-injuries, 88 A-injuries, 357 B-injuries, 309 C-injuries, and 2,622 occupants without injuries. It is important to note that the percentage for each attribute was calculated by dividing the number of crashes by the total crashes in Wisconsin and tribal lands, respectively. The KAB and KA crash rates were determined by dividing the number of KAB and KA crashes by the total number of crashes for the corresponding attributes.

In the case of crashes on tribal lands, 3,211 (94.6%) crashes involved 1,865 (58.1%) male drivers and 1,346 (39.6%) female drivers, resulting in a difference of 519 more males than females involved in these incidents.

Regarding age groups, both statewide crashes and crashes on tribal lands exhibited similar rates. The only noticeable difference was in the age range of 45 to 64, which had a 3.1% higher total percentage in tribal lands compared to statewide data. For KAB crashes, the age groups of 15 to 24 years and 25 to 44 years had rates that were 3.1% and 4.1% higher in tribal lands than statewide, respectively. Based on these observations, it can be inferred that individuals under the age of 44 had a greater likelihood of being involved in fatal and serious injury crashes on tribal lands compared to the rest of the state.

In addition to sex and age, **Table 2** presents key factors related to general crash analysis, such as speeding, impaired driving, pedestrian involvement, hit-and-run incidents, and safety belt usage. Specifically, there were 316 (9.3%) crashes associated with impaired driving on tribal lands, a rate that is 3.3% higher than in statewide crashes. Furthermore, all key factors exhibited higher KAB and KA rates on tribal lands compared to statewide data, particularly in impaired driving and pedestrian-involved crashes, which were 10.5% and 9.3% higher, respectively.

Table 2. Summary Statistics for Severity level of Crashes in Wisconsin from 2017 to 2021.

| Variable | Occupant Injuries by Severity Level | |
|---|---|---|
| | Statewide Crashes | Crashes in Tribal land |

|  | Total | % | KAB | % | KA | % | Total | % | KAB | % | KA | % |
|---|---|---|---|---|---|---|---|---|---|---|---|---|
| Sex |  |  |  |  |  |  |  |  |  |  |  |  |
| Female | 246957 | 36.7 | 27971 | 11.3 | 4551 | 1.8 | 1346 | 39.6 | 170 | 12.6 | 32 | 2.4 |
| Male | 358933 | 53.4 | 46668 | 13.0 | 11423 | 3.2 | 1865 | 54.9 | 287 | 15.4 | 71 | 3.8 |
| Age Group |  |  |  |  |  |  |  |  |  |  |  |  |
| ≤ 4 | 33 | 0.0 | 10 | 30.3 | 4 | 12.1 | 0 | 0.0 | 0 | 0.0 | 0 | 0.0 |
| ≥ 5 and ≤ 14 | 795 | 0.1 | 311 | 39.1 | 68 | 8.6 | 4 | 0.1 | 2 | 50.0 | 1 | 25.0 |
| ≥ 15 and ≤ 24 | 155109 | 23.1 | 19958 | 12.9 | 3658 | 2.4 | 756 | 22.3 | 121 | 16.0 | 24 | 3.2 |
| ≥ 25 and ≤ 44 | 219105 | 32.6 | 27250 | 12.4 | 6124 | 2.8 | 1117 | 32.9 | 184 | 16.5 | 42 | 3.8 |
| ≥ 45 and ≤ 64 | 157308 | 23.4 | 17912 | 11.4 | 4176 | 2.7 | 899 | 26.5 | 98 | 10.9 | 25 | 2.8 |
| ≥ 55 and ≤ 74 | 44596 | 6.6 | 5273 | 11.8 | 1153 | 2.6 | 258 | 7.6 | 31 | 12.0 | 4 | 1.6 |
| ≥ 75 | 28041 | 4.2 | 3847 | 13.7 | 776 | 2.8 | 170 | 5.0 | 21 | 12.4 | 7 | 4.1 |
| Key Elements |  |  |  |  |  |  |  |  |  |  |  |  |
| Speeding | 94657 | 14.1 | 17753 | 18.8 | 4896 | 5.2 | 484 | 14.3 | 108 | 22.3 | 33 | 6.8 |
| Impaired | 40445 | 6.0 | 12795 | 31.6 | 5002 | 12.4 | 316 | 9.3 | 133 | 42.1 | 56 | 17.7 |
| Pedestrian | 6908 | 1.0 | 4752 | 68.8 | 1592 | 23.0 | 32 | 0.9 | 25 | 78.1 | 12 | 37.5 |
| Hit & Run | 97800 | 14.5 | 5755 | 5.9 | 1760 | 1.8 | 336 | 9.9 | 21 | 6.3 | 9 | 2.7 |
| Safety Belt | 50327 | 7.5 | 10005 | 19.9 | 5755 | 11.4 | 288 | 8.5 | 66 | 22.9 | 40 | 13.9 |
| Grand Total | 672363 |  | 77516 | 11.5 | 16450 | 2.4 | 3396 |  | 465 | 13.7 | 108 | 3.2 |

The comparisons reveal that similar crash types on tribal lands yield higher KAB and KA severity rates compared to other regions within the state of Wisconsin. Given these observations and findings, it is crucial to develop a tool capable of systematically identifying and analyzing high-risk areas in tribal lands to aid in the formulation of potential safety intervention strategies. This paper proposes a dashboard-based safety system that displays the potential elements capable of achieving the aforementioned objective using the DT4000 database. It is important to note that due to challenges in data acquisition, variations in road facilities, and differing road environments, we have only utilized data from Wisconsin for a unified analysis and have refrained from making horizontal comparisons with data from other states.

To enhance the understanding of the dashboard concept, **Figure 2** illustrates the ongoing Wisconsin tribal dashboard prototype created using PowerBI (Ferrari & Russo, 2016). The dashboard consists of various modules that offer map visualization, basic crash statistics, and crash count ranking. The map visualization module presents crash points with varying severity levels in different areas, making it advantageous for identifying high-risk regions in tribal lands. The basic crash statistics module provides an intuitive overview of the five-year tribal crashes, displaying the total number of crashes, facilities, and injuries, along with the crash counts categorized by different severity levels. The crash counts ranking module offers a histogram of the ranking of crash counts by tribes and a bar chart of the top 10 crash types. These comparative showcases can serve as a good starting point for in-depth crash analysis. All the modules are dynamic and update with new data. Additionally, the dashboard capability introduces various potential dashboard element analyses, which are discussed in the following section.

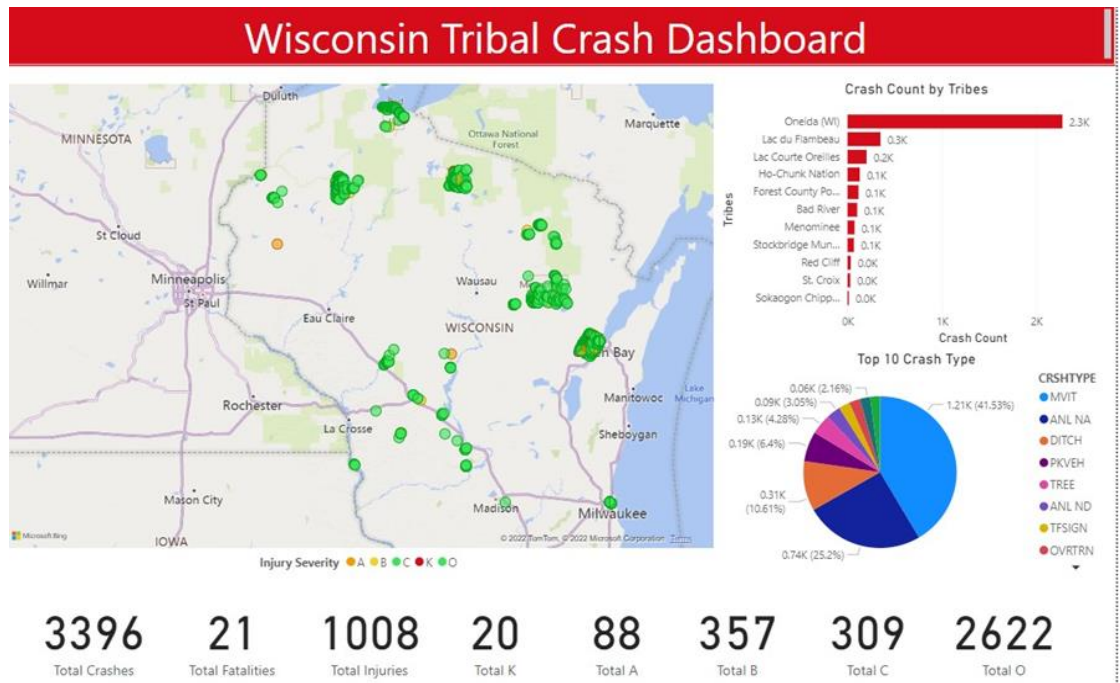

Figure 2. Wisconsin tribal crash dashboard prototype.

**4. Potential tribal dashboard elements analysis**

A primary objective of the dashboard is to pinpoint high-risk areas using intuitive statistics and methods. Although numerous studies (e.g., Davis, 2014; Srinivasan & Bauer, 2013) have put forth prevailing techniques for identifying roads with elevated crash frequency and severity, their models are not suitable for this study. For instance, the safety performance function (Srinivasan & Bauer, 2013) employs crash attributes (e.g., injury severity), road geometrics (e.g., lane width), and traffic conditions (e.g., annual average daily traffic (AADT)) to determine sites with the greatest potential for safety interventions. However, the current dataset lacks AADT data for roads within tribal lands. Additionally, the dataset does not provide detailed information on BIA roads, instead categorizing them as local road types, rendering the analysis of specific BIA roads unfeasible. A novel and comprehensive approach is required to locate areas with high crash severity within tribal lands, utilizing the available data elements and attributes from the Wisconsin DT4000 crash report.

*4.1. Location based tribal crash severity analysis*

Given the aforementioned limitations, this study refines the target area by utilizing region and roadway function elements, categorized as follows: (i) urban and rural areas and (ii) highway and non-highway roads. Specifically, highways are defined as state trunk highways (STH), U.S. highways (USH), and interstate highways (IH), while non-highways encompass all other roads, such as county trunk highways (CTH) and local roads. It should be noted that we combined CTH and local roads into the non-highway category for simplicity in this analysis, as separating them proved to be less meaningful after the investigation. **Table 3** presents the summary statistics of different road types for severity levels in both state and tribal lands.

Table 3. Summary of Road Types of Statistics for Severity Level in Wisconsin from 2017 to 2021.

| Statewide Crashes | Total | KAB | % | KA | % |
|---|---|---|---|---|---|
| Total Crashes | 672363 | 77516 | 11.5 | 16450 | 2.4 |
| Highway (STH USH IH) | 257204 | 30962 | 12.0 | 6843 | 2.7 |
| Non-highway (CTH Local) | 415159 | 46554 | 11.2 | 9607 | 2.3 |
| Rural Highway | 138268 | 17061 | 12.3 | 4460 | 3.2 |
| Rural Non-highway | 147508 | 18517 | 12.6 | 5064 | 3.4 |
| Urban Highway | 118936 | 13901 | 11.7 | 2383 | 2.0 |
| Urban Non-highway | 267651 | 28037 | 10.5 | 4543 | 1.7 |
| Crashes in Tribal land | | | | | |
| Total Crashes | 3396 | 465 | 13.7 | 108 | 3.2 |
| Highway | 1360 | 185 | 13.6 | 37 | 2.7 |
| Non-highway | 2036 | 280 | 13.8 | 72 | 3.5 |
| Rural Highway | 543 | 87 | 16.0 | 27 | 5.0 |
| Rural Non-highway | 1040 | 153 | 14.7 | 51 | 4.9 |
| Urban Highway | 817 | 98 | 12.0 | 10 | 1.2 |

| | | | | | | |
|---|---|---|---|---|---|---|
| Urban Non-highway | 996 | 127 | 12.8 | 20 | 2.0 | |

**Table 3** illustrates that the overall KAB and KA crash rates in tribal lands were higher than the statewide crash rate, with increases of 2.2% and 0.8%, respectively. This trend was particularly noticeable in rural areas. Specifically, for rural highway segments, there was a 3.7% and 2.1% increase in KAB and KA rates, respectively, compared to statewide statistics. For rural non-highway segments, there were 2.1% and 1.5% increases in KAB and KA rates, respectively. Therefore, we can conclude that the rural regions within tribal lands exhibit higher severity rates, particularly on rural highways. Since BIA roads in Wisconsin are classified as non-highway elements, it can be inferred that BIA roads in Wisconsin have a higher crash severity rate than other road segments.

A comprehensive analysis of crash severity rates for each tribe has been conducted. The dataset and cartographic boundaries of the tribal lands have been imported into Esri ArcMap. **Table 4** displays the detailed KAB and KA rates for each tribe, along with their corresponding rankings. The top three most dangerous tribes based on the KAB rate are the Menominee Indian Tribe, Lac Courte Oreilles Band, and St. Croix Chippewa Indians, which exhibit KAB rates 7% higher than the average.

Table 4. Summary of KAB/KA Crash Rate for Each Tribe in Wisconsin.

| Tribes Name | Total | KAB | KAB Rate | KA | KA Rate | KAB Ranking | KA Ranking |
|---|---|---|---|---|---|---|---|
| Menominee Indian Tribe | 74 | 16 | 21.62% | 6 | 8.11% | 1 | 3 |
| Lac Courte Oreilles Band | 202 | 42 | 20.79% | 18 | 8.91% | 2 | 2 |
| St. Croix Chippewa Indians | 29 | 6 | 20.69% | 4 | 13.79% | 3 | 1 |
| Bad River Band | 103 | 20 | 19.42% | 6 | 5.83% | 4 | 6 |

| | | | | | | |
|---|---|---|---|---|---|---|
| Lac Du Flambeau Band | 349 | 58 | 16.62% | 14 | 4.01% | 5 | 7 |
| Sokaogon Chippewa Community | 14 | 2 | 14.29% | 0 | 0.00% | 6 | 11 |
| Ho-Chunk Nation | 130 | 17 | 13.08% | 5 | 3.85% | 7 | 8 |
| Oneida Tribe Of Indians | 2277 | 287 | 12.60% | 47 | 2.06% | 8 | 9 |
| Red Cliff | 34 | 4 | 11.76% | 2 | 5.88% | 9 | 4 |
| Stockbridge-Munsee Community | 68 | 6 | 8.82% | 4 | 5.88% | 10 | 5 |
| Forest County Potawatomi Community | 116 | 7 | 6.03% | 2 | 1.72% | 11 | 10 |

Map-based representation and hot-spot analysis are two crucial components of the dashboard. Two case studies, one with a high KAB rate (Lac Courte Oreilles) and another with a low KAB rate (Oneida), are demonstrated in **Figure 3**. Specifically, **Figure 3(a)** displays all crashes in the Lac Courte Oreilles Band, symbolized by color according to their severity level. Utilizing the spatial statistics function in ArcMap, it is evident that most crashes occurred in the northwest part of the reservation and were typically located on CTH (highlighted in the red box). This observation confirms that rural highways are more dangerous. In the Oneida case, most crashes were situated on urban highways (highlighted in the red box). Given Oneida's proximity to urban areas, it can be inferred that improved infrastructure and lower speed limits effectively reduce high-severity crash rates. It should be noted that there are various ways to generate crash heat maps; in this case, we use the ArcMap function as a demonstration. Overall, this mapping analysis offers a reasonable example of how the dashboard system can effectively locate road segments for potential safety interventions. This type of display could be automated and incorporated into the dashboard.

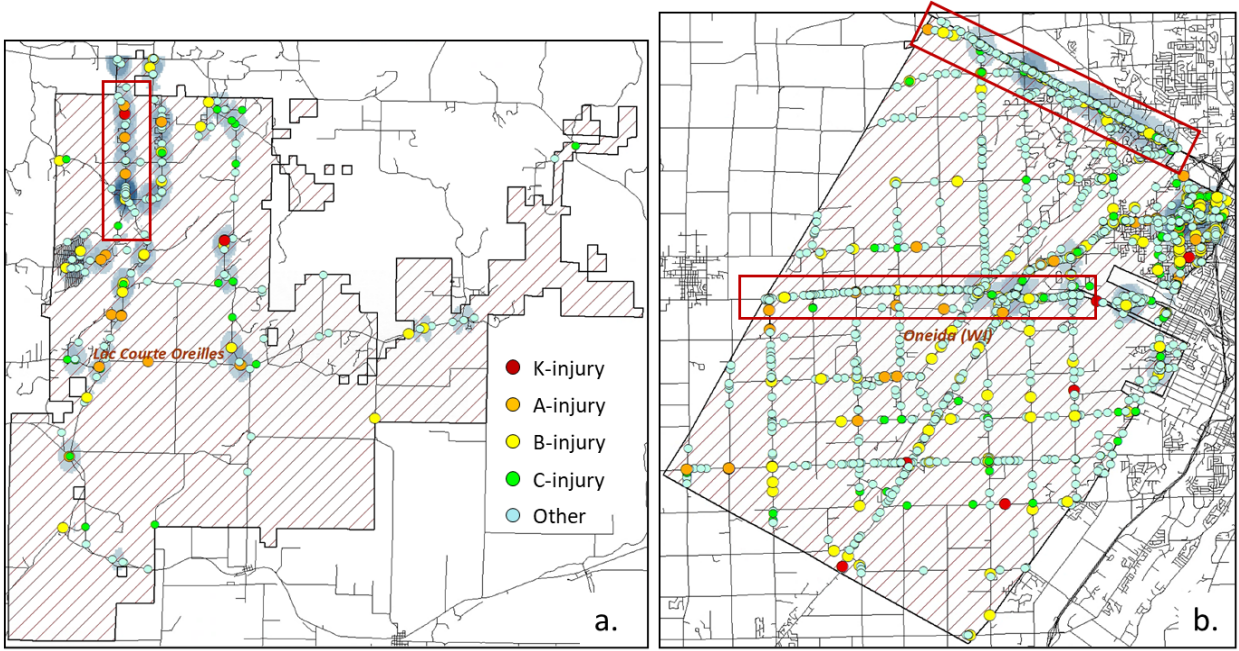

Figure 3. Case study in a) Lac Courte Oreilles Band and b) Oneida.

## *4.2. Tribal crash type analysis*

The crash type analysis represents another potential element to be incorporated into the safety dashboard. **Figure 3** summarizes the top 10 crash types on tribal lands, ranked by total crashes and KAB crashes, along with a corresponding comparison to statewide statistics.

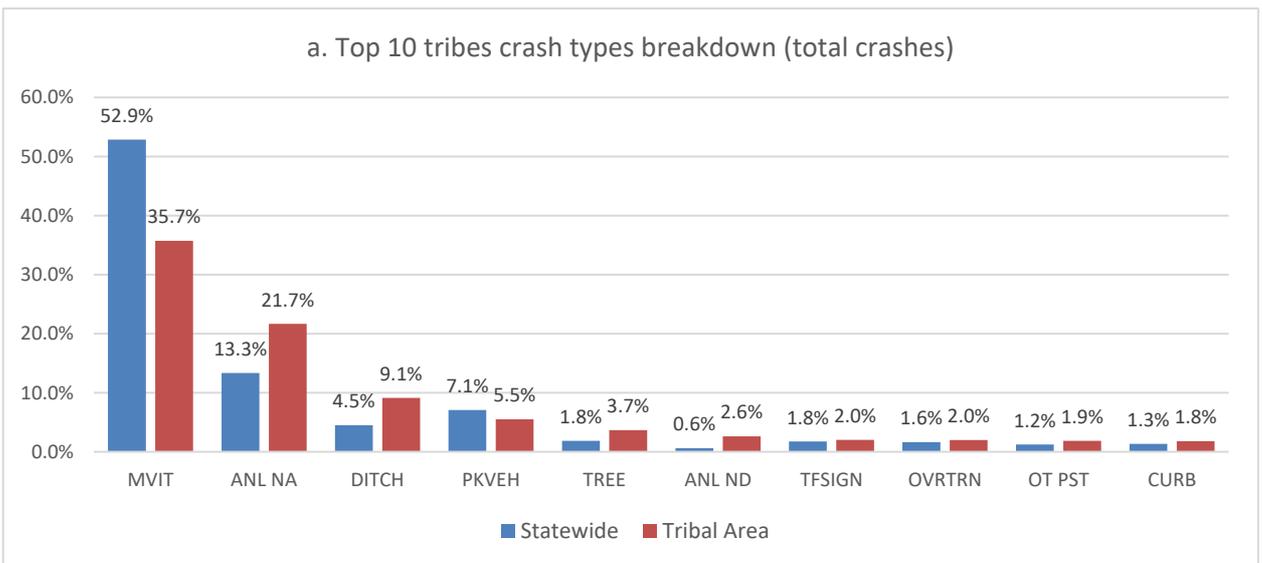

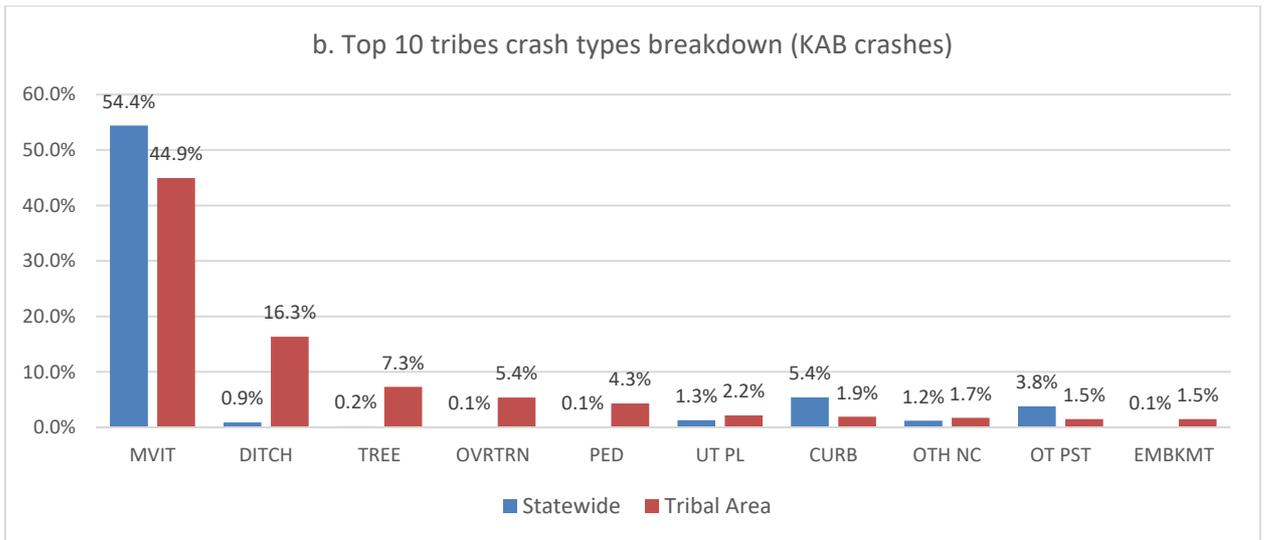

Figure 4. Top 10 tribal crash types compared with statewide statistics in Wisconsin based on a) total crashes, b) KAB crashes.

It has been observed that animal non-domestic alive (ANL NA), animal non-domestic dead (ANL ND), ditch, and tree crashes occurred more frequently in tribal lands compared to statewide crashes, as depicted in **Figure 4(a)**. Specifically, the percentages of the ditch and tree types were twice as high as the statewide figures. For KAB crashes shown in **Figure 4(b)**, the gaps between tribal lands and statewide numbers for ditch and tree types were even more significant. Meanwhile, the differences for overturning, pedestrian, and embankment types are relatively higher compared to the gaps in total cases.

With the above analysis and observations, the dashboard can serve as a starting point for secondary analysis. For instance, an investigation could examine the relationship between notable crash types and driver factors, such as impaired driving, speeding, occupant restraint use within specific tribal lands, or roadway geometric characteristics for particular segments with high ditch or embankment crash type rates. The safety dashboard would enable users to interactively delve deeper by applying additional filters.

## 5. Conclusions

Effective data-driven approaches for safety analysis are crucial in tribal lands due to their high crash rates and severities. With eleven tribal nations situated within the State of Wisconsin, there is a strong and enduring interest in analyzing Wisconsin tribal crashes. However, the current Wisconsin crash database (i.e., DT4000) has certain limitations for further safety performance analysis, including a lack of sufficient traffic volume data, AADT data, and detailed information on BIA roads. To overcome these limitations and conduct feasible, reproducible, and effective tribal crash analysis, this paper presents a generic dashboard of crash analysis techniques and demonstrates its application to the current Wisconsin tribal crash data.

The dashboard connects to the DT4000 database, encompassing a comprehensive record of reported crashes in Wisconsin, integrating crash-level, vehicle-level, and person-level elements and attributes. To obtain an intuitive understanding and analysis of tribal crashes, the dashboard identifies high-risk tribal lands by examining crash severity and comparing Wisconsin statewide crash statistics and tribal averages. The findings indicate that roads in tribal lands are more hazardous than those in other areas of Wisconsin, with the KA/KAB rate being significantly higher in rural tribal regions. Analyzing the crash severity rate for each tribe in Wisconsin aids in pinpointing locations for safety resource allocation (e.g., guardrail installation or targeted enforcement activities). Furthermore, an investigation of crash types provides insights into the causes behind tribal crashes, potentially assisting agencies in identifying and addressing underlying issues.

In summary, the proposed tribal safety dashboard enables a more intuitive understanding of tribal crashes from a statistical standpoint, helping to harness available data sources and create effective safety interventions. The dashboard can be updated as new data is fed from the database, providing flexibility and reproducibility. Various in-depth crash analyses can be conducted using the dashboard in the future. The dashboard has the potential to analyze tribal crash trends based on historical data and make crash predictions for future periods. Other data tracking or monitoring methods can also be employed to observe critical indicators of tribal crash data. These analytical measures could better assist agencies in taking proactive steps to prevent safety issues. It is important to note that the dashboard is a flexible concept that can be enhanced as more data becomes available. Additional indicators can be incorporated based on a specific state's database. For example, crash density and crash rate could be integrated into the dashboard for comparison and analysis if data is available. From a practical perspective, these analytical functions and indicators can be combined into an online dashboard tool to facilitate applications. Leveraging the comprehensive analysis results from the dashboard, states can implement more informed measures to reduce tribal crashes and improve tribal safety. In future work, a working prototype will be developed based on the conceptual framework presented in this paper.

**Funding details**

Development of the Wisconsin Crash Database and DT4000 crash report was sponsored by the Wisconsin Department of Transportation. The work was supported by the Traffic Operations and Safety (TOPS) Laboratory.

**Disclosure statement**

The authors report there are no competing interests to declare.